\documentclass[pre,twocolumn,floatfix]{revtex4}
\usepackage{amsmath}
\usepackage{amssymb}
\usepackage{epsfig}

\begin{document}

% Title must be 150 words or less
\title{Mesoscopic organization reveals the constraints governing
Caenorhabditis elegans nervous system}
% Insert Author names, affiliations and corresponding author email.

\author{Raj Kumar Pan}
\email{rajkp@imsc.res.in}
\affiliation{%
The Institute of Mathematical Sciences, C.I.T. Campus, Taramani,
Chennai - 600 113 India
}%
\author{Nivedita Chatterjee}
\email{drnc@snmail.org}
\affiliation{%
Vision Research Foundation, Sankara Nethralaya, 18 College Road,
Chennai - 600 006 India
}%
\author{Sitabhra Sinha}
\email{sitabhra@imsc.res.in}
\affiliation{%
The Institute of Mathematical Sciences, C.I.T. Campus, Taramani,
Chennai - 600 113 India
}%

\date{\today}

% Please keep the abstract between 250 and 300 words
\begin{abstract}
One of the biggest challenges in biology is to understand how activity at
the cellular level of neurons, as a result of their mutual interactions,
leads to the observed behavior of an organism responding to a variety of
environmental stimuli. Investigating the intermediate or mesoscopic level
of organization in the nervous system is a vital step towards understanding
how the integration of micro-level dynamics results in macro-level
functioning. The coordination of many different co-occurring processes at
this level underlies the command and control of overall network activity.
In this paper, we have considered the somatic nervous system of the
nematode Caenorhabditis elegans, for which the entire neuronal connectivity
diagram is known. We focus on the organization of the system into modules,
i.e., neuronal groups having relatively higher connection density compared
to that of the overall network. We show that this mesoscopic feature cannot
be explained exclusively in terms of considerations such as, optimizing for
resource constraints (viz., total wiring cost) and communication efficiency
(i.e., network path length). Even including information about the genetic
relatedness of the cells cannot account for the observed modular structure.
Comparison with other complex networks designed for efficient transport (of
signals or resources) implies that neuronal networks form a distinct class.
This suggests that the principal function of the network, viz., processing
of sensory information resulting in appropriate motor response, may be
playing a vital role in determining the connection topology. Using modular
spectral analysis we make explicit the intimate relation between function
and structure in the nervous system. This is further brought out by
identifying functionally critical neurons purely on the basis of patterns
of intra- and inter-modular connections.  Our study reveals how the design
of the nervous system reflects several constraints, including its key
functional role as a processor of information.
\end{abstract}

\pacs{89.75.Hc,89.75.Fb}
% PACS, the Physics and Astronomy Classification Scheme.
%89.75.Hc Networks and genealogical trees
%05.45.-a Dynamical systems nonlinear, 
%89.75.-k Complex systems, 
%05.65.+b Criticality, self-organized 
%89.75.Fb Structures and organization in complex systems

\maketitle

% Please keep the Author Summary between 150 and 200 words
% Use first person. PLoS ONE authors please skip this step. 
% Author Summary not valid for PLoS ONE submissions.   
\section{Author Summary}
Why are brains wired in the manner that is observed in nature? To address
this, we analyze the structure of the Caenorhabditis elegans neuronal
network at an intermediate level of its organization into groups of densely
connected neurons. These structural units, called modules, are composed of
multiple neuronal types which rules out the possibility that this is a
straightforward partition of the network into sensory and motor regions.
Nor do they have a simple anatomical interpretation (e.g., in terms of
ganglia). We investigate the possibility that this modular structure
developed through an evolutionary drive to optimize a nervous system
subject to different constraints. We show that neither wiring cost
minimization nor maximization of communication efficiency are adequate in
explaining the observed patterns in the network topology. The distinction
with other complex networks designed for efficient signal propagation (such
as the Internet) suggests that the neuronal network's key functional
requirement of information processing may play a crucial role in
determining its overall structure. Analyzing the network organization at
the mesoscopic level also allows us to identify neurons belonging to known
functional circuits and predict those that may play a vital part in some as
yet undetermined function.

\section{Introduction}
The relatively simple nervous systems of invertebrate organisms provide
vital insights into how nerve cells integrate sensory information from the
environment, resulting in a coordinated response. Analysing the
intermediate or mesoscopic level of organization in such systems is a
crucial step in understanding how micro-level activity of single neurons
and their interactions eventually result in macro-level behavior of the
organism~\cite{Tononi94}.  The nematode {\em Caenorhabditis elegans} is a
model organism on which such an analysis can be performed, as its entire
neuronal wiring layout has been completely mapped~\cite{White86}.  This
information enables one to trace in full the course of activity along the
neuronal network, from sensory stimulation to motor
response~\cite{CelegansII}.  We study its somatic nervous system,
comprising 282 neurons that control all activity except the pharyngeal
movements. This can lead to an understanding of the command and control
processes occurring at the mesoscopic level that produce specific
functional responses, including avoidance behavior and movement along a
chemical gradient. The neuron locations as well as their connections being
completely determined by the genetic program, are invariant across
individual organisms~\cite{CelegansII}. Further, unlike in higher organisms,
the connections do not change with time in the adult
nematode~\cite{Schnabel97}.  In combination with the possibility of
experimenting on the role of single neurons in different functional
modalities, these invariances allow one to uniquely identify the important
neurons in the system having specific behavioral tasks.

The recent developments in the theory of complex graphs has made available
many analytical tools for studying biological
networks~\cite{Newman03,Albert02}. The initial emphasis was on developing
gross macroscopic descriptions of such systems using measures such as
average path length between nodes of the network, the clustering among
nodes and the degree sequences (the number of links per node). However,
such global characterizations of systems ignore significant local
variations in the connection topology that are often functionally
important.  Therefore, investigating the network at a mesoscopic level
which consider the broad patterns in the inhomogeneous distribution of
connections, may reveal vital clues about the working of an organism that
could be hidden in a global analysis.  Further, these large-scale features
help in understanding how coordination and integration occurs across
different parts of the system, in contrast to a study of microscopic
patterns comprising only a few neurons, e.g., motifs~\cite{Reigl04}.

The existence of {\em modules}, marked by the occurrence of groups of
densely connected nodes with relatively fewer connections between these
groups~\cite{Girvan02}, provides a natural meso-level description of many
complex systems~\cite{Hartwell99}.  In biological networks, such modular
organization has been observed across many length scales, from the
intra-cellular protein-protein interaction
system~\cite{Schwikowski00,Rives03} to food webs comprising various
species populations~\cite{Krause03}. The relation between certain modules
and specific functions, e.g., in metabolic networks~\cite{Guimera05}, helps
us to understand how different functions are coordinated in the integral
performance of complex biological networks.

\begin{figure}[!ht]
\begin{center}
\includegraphics[width=0.9\linewidth]{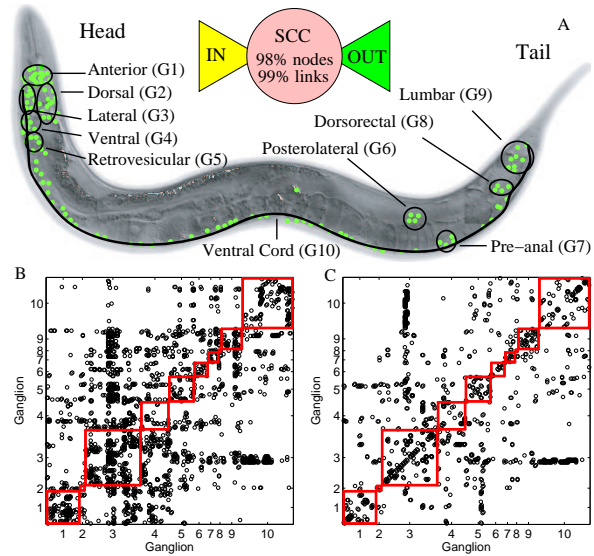}
\end{center}
	\caption{
	{\bf Neuronal position and connectivity in the somatic nervous system of
	the nematode C. elegans indicating the different ganglia.} (a) Schematic
	diagram of C. elegans, indicating the different ganglia.  (Inset)
	Schematic representation of connectivity between the neurons, partitioned
	into a {\em strongly connected component} (SCC), an {\em in-component}
	(IN), and an {\em out-component} (OUT). A directed path exists from any
	neuron in IN to any neuron in OUT through neurons in SCC, all of whose
	members can be reached from each other. The large SCC suggests that it is
	possible to transfer signals between almost all neurons of the network.
	The IN and OUT components have only $1.5 \%$ and $0.5 \%$, respectively,
	of the 279 connected neurons in the somatic nervous system.  (b, c) The
	connectivity matrix corresponding to the (b) Synaptic and (c)
	Gap-junctional connections between the somatic system neurons.  In all
	figures, the partition symbols correspond to (G1) Anterior, (G2) Dorsal,
	(G3) Lateral, (G4) Ventral, (G5) Retrovesicular, (G6) Posterolateral,
	(G7) Preanal, (G8) Dorsorectal and (G9) Lumbar ganglion, and (G10) the
	Ventral cord.}
\label{fig:celegans}
\end{figure}

Modular organization in the brains of different species have been observed,
both in {\em functional} networks derived from EEG/MEG and fMRI experiments
and in {\em structural} networks obtained from tracing anatomical
connections~\cite{Bullmore09}. Functionally defined networks are
constructed by considering brain areas, comprising a large number of
localized neuronal groups, which are linked if they are simultaneously
active. Such systems have been shown to be modular for both
human~\cite{Ferrarini08,Meunier09} and non-human~\cite{Schwarz08} subjects.
Tract-tracing studies in the brains of cat~\cite{Scannell95} and
macaque~\cite{Young93} have also revealed a modular layout in the
structural inter-connections between different cortical areas. However, as
neurons are the essential building blocks of the nervous system, ideally
one would like to explore the network of interconnections between these
most basic elements.  In the extremely complicated mammalian brains, it is
so far only possible to analyze such networks for extremely limited regions
that do not give a picture of how the system behaves as a
whole~\cite{Humphries06}.  The relative simplicity of the nervous system of
C. elegans allows a detailed analysis of the network, defined in terms of
both electrical (gap junctional) and chemical (synaptic) connections
between the neurons (Fig.~\ref{fig:celegans}).

The ubiquity of modularity in brain networks leads to the obvious question
about how to explain the evolution of such a structural
organization~\cite{Pan07c}. One possible reason for the existence of
modular architecture is that it may result in low average path length
(which is associated with high efficiency of signal communication) and high
clustering (that allows local segregation of information processing) in
networks~\cite{Pan09a}. An alternative possibility is that segregation of
neurons into spatially localized communities minimizes the total cost
associated with the wiring length (the physical distance spanned by
connections between neurons). This cost arises from resources associated
with factors such as wiring volume as well as metabolism required for
maintenance and propagation of signals across long distances~\cite{Chen06}.
Developmental constraints, such as the lineage relations between different
neurons may also play an important role in determining the connection
topology of the neuronal network~\cite{Deppe78}. In addition, the existence of
empirically determined circuits responsible for specific functions (such
as, movement associated with exploratory behavior, egg laying, etc.) in the
C. elegans nervous system, raises the intriguing possibility that
structurally defined modules are associated with definite functional
roles~\cite{Arenas08a}. The invariant neuronal connectivity profile of C.
elegans allows us to explore the contributions of the above mentioned
structural, developmental and functional constraints in governing the
mesoscopic organization of the nervous system.

In this paper, we begin our analysis of the organization of the C. elegans
nervous system by identifying structurally defined modules in the network
of neurons linked by synapses and gap-junctions. Next, we investigate
whether the observed modular structure can be explained by using arguments
based on universal principles. Such criteria, which include minimizing the
cost associated with neuronal connections~\cite{Chen06,Escudero07} and
their genetic encoding~\cite{Itzkovitz08}, or, decreasing the signal
propagation path~\cite{Kaiser06,Ahn06}, have recently been proposed to
explain observed patterns of neuronal position and connectivity. 
Complementing the above studies, we seek to understand the factors 
determining the topological arrangement of the nervous system, given
the physical locations of the neurons.
We determine the role of physical proximity between a pair of neurons in
deciding the connection structure, by investigating the correlation between
their spatial positions and their modular membership.  We also compare
these modules with the existing classification of the nematode nervous
system into several ganglia, as the latter have been differentiated in
terms of anatomical localization of their constituent neurons. Results
from the above analysis suggests that resource constraints such as wiring
cost cannot be the sole deciding factor governing the observed meso-level
organization. We also show that the modules cannot be only a result of the
common lineage of their member nodes.  

It is natural to expect that the structure of the nervous system is
optimized to rapidly process signals from the environment so that the
organism can take appropriate action for its survival~\cite{Brenner00}. By
looking at the deviation between the actual network and a system optimized
for maximal communication efficiency in conjunction with minimum wiring
cost, we infer the existence of additional functional constraints possibly
related
to processing of information (i.e., other than rapid signal transmission).
Information processing refers to the transformation of signals~\cite{Alon07}, 
such as
selective suppression of activation in specific pathways, which allows
different sensory stimuli to initiate response in distinct sets of motor 
neurons. Absence of such transformations would result in 
non-specific
arousal of a large number of connected neurons, and will not achieve
the high level of control and coordination necessary for performing a
variety of specific functional tasks.
This is also related to the observation of relatively high clustering
in C. elegans neuronal network as compared to other information networks
(e.g., electronic logic circuits~\cite{Cancho01}).  Previous investigation
of the role of clustering in the performance of neural network models in
associative recall of stored patterns, has shown that lower clustering
exhibits much better performance~\cite{Kim04}.  
It has also been shown that the clustering in C. elegans neuronal 
network is higher than that in degree-conserved randomized networks having the 
same wiring cost~\cite{Ahn06}. Therefore, the presence of enhanced clustering
in a system that has evolved under intense competition for
survival may imply it plays a key role in processing
information. 

This brings us to the possibility that the observed distribution of neurons
among modules is closely related to the behavioral requirements of the
organism. For this purpose, we investigate the relation between the modules
and the different functional circuits governing specific behavioral aspects
of C. elegans. We identify the functional importance of certain key neurons
by observing their relative connectivity within their module as compared to
that with neurons belonging to other modules.  By looking at the
correlation between local and global connectivity profiles of individual
neurons, we observe that the nematode nervous system is 
different from systems designed for rapid signal transfer, including other
information networks occurring in the technological domain, such as the
Internet.  Further, in contrast to previous observations on the similarity
between biological signalling networks having different
origins~\cite{Milo04,Klemm05}, we find that the C. elegans neuronal network
has properties distinct from at least one other biological network
involved in signalling, the protein interaction
network (viz., in terms of the assortativity and the
role-to-role connectivity profile)~\cite{Maslov02,Guimera07}.

Thus, the analysis of the network at the mesoscopic level provides an
appropriate framework for identifying the roles that different classes of
constraints (developmental, structural and functional) play in determining
the organization of a nervous system. It allows us to infer the existence
of criteria related to processing of information governing the observed
modular architecture in C. elegans neuronal inter-connections. It also
provides the means for identifying neurons having key roles in the
behavioral performance of the organism exclusively from anatomical
information about their structural connectivity. Our results can help
experimentalists in focusing their attention to a select group of neurons
which may play a vital part in as yet undetermined functions. 

% Results and Discussion can be combined.
\section{Results}
\subsection{Modular structure of the C.elegans somatic nervous system}
We begin our study of the mesoscopic organization of the network by
focusing on identifying its modular arrangement.  In order to determine the
community structure of the C. elegans neuronal network, we perform an
optimal partitioning of the system into modules, that corresponds to the
maximum value of modularity parameter, $Q$ (see Methods for details on
modularity measure $Q$ and the algorithm used for modularity
determination).  We have considered different cases corresponding to the
different types of neuronal connections (viz., gap junctions and synapses)
and the nature of such connections (i.e., unweighted or weighted by the
number of connections between a given pair).  While the gap junctional
network is undirected, the directional nature of signal propagation through
a synapse implies that the synaptic network is directed.  For each network,
we have obtained using the spectral method, 
the maximum modularity $Q_M$ and the corresponding number
of partitions (Table~\ref{tab:mod_network}).  Note that, the number of
modules and their composition is dependent on the type of connections we
consider.  

\begin{table*}[!ht]
	\caption{ 
	{\bf Modularity of the C. elegans neuronal network.}}
\begin{scriptsize}
	\begin{tabular}{|l|c|c|c|c|c|c|c|c|c|c|c|c|} \hline
	Network & 
	\multicolumn{6}{c|} {Un weighted} & 
	\multicolumn{6}{c|} {Weighted}\\\cline{2-13}
	& $Q_g$ & $Q_M$ & $m_M$ & $I$ & $Q_M^{\rm rand}$ & $m_M^{\rm rand}$ & $Q_g$ & $Q_{M}$ & $m_M$ & $I$ & $Q_M^{\rm rand}$ & $m_M^{\rm rand}$ \\\hline
 Gap Jn  &0.207& 0.630 & 11 & 0.326 & 0.467 $\pm$ 0.010 & 10.5 $\pm$ 2.2 & 0.170 & 0.657 & 15 & 0.347 & 0.519 $\pm$ 0.022 & 12.5 $\pm$ 3.9 \\\hline
 Synaptic&0.149& 0.349 & 2  & 0.257 & 0.192 $\pm$ 0.013 & 3.6 $\pm$ 0.7 & 0.211 & 0.472 & 4  & 0.314 & 0.307 $\pm$ 0.018 & 5.6 $\pm$ 1.4 \\\hline
 Combined&0.169& 0.378 & 3  & 0.306 & 0.156 $\pm$ 0.012 & 3.2 $\pm$ 0.7 & 0.203 & 0.491 & 6  & 0.376 & 0.258 $\pm$ 0.015 & 4.9 $\pm$ 1.5 \\\hline
 \end{tabular}
\end{scriptsize}
\begin{flushleft}
The modularity of the network is measured using the parameter $Q$, which
requires a knowledge of the partitions or communities which divide the
network. We obtain the modularity measure, $Q_g$, on assuming the
communities to correspond to the ganglia. Its positive values indicate that
neurons in the same ganglion have high density of inter-connections. We
have also obtained $Q$ by determining the modules of the network using a
spectral method, the corresponding values being indicated by $Q_{M}$. The
relatively high values of $Q_M$ compared to $Q_g$, indicates that the
ganglia do not match with this optimal partitioning of the network. The
measures, $Q_g$ and $Q_{M}$, as well as the number of modules, $n_M$, have
been obtained for both unweighted and weighted networks consisting of
either gap junctions or synapses or both. We calculate the overlap between
the ganglionic and the optimal partition of the network using the
normalized mutual information index, $I$. For the case of perfect match
between the two, the index, $I=1$, whereas if they are independent of each
other, $I=0$. The measured values of $I$ indicate that the overlap between
the different modules and the anatomically defined ganglia is not high.
The modular nature of the somatic nervous system is
emphasised by comparing the empirical network with networks obtained by
randomizing the connections, keeping the degree of each neuron fixed. The
mean and standard deviation of the modularity $Q_M^{\rm rand}$ and the
corresponding number of partitions $m_M^{\rm rand}$ are shown for both
weighted and unweighted networks, and for the different types of
connections. For all cases, the randomized networks show a significantly
lower modularity than the empirical network.
\end{flushleft}
 \label{tab:mod_network}
\end{table*}

\begin{figure}[!ht]
\begin{center}
\includegraphics[width=0.95\linewidth]{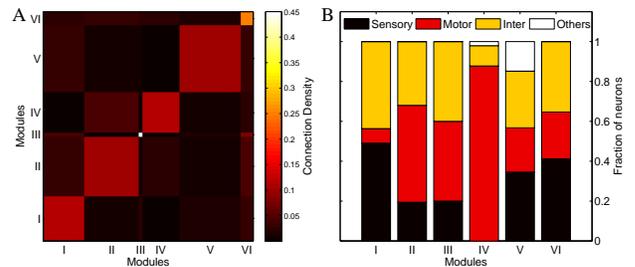}
  \end{center}
	\caption{
	{\bf Modular interconnectivity and decomposition according to neuron
	type.} (a) Matrix representing the average connection density between
	neurons occurring within modules and those in different modules. The
	figure indicates that neurons within a module are densely interconnected
	compared to the overall connectivity in the network.  (b) The modules are
	decomposed according to the different neuron types comprising them. The
	figure shows that the modules are not simply composed of a single type of
	neuron.}
\label{fig:modular_connectivity}
\end{figure}

As we want to consider all connections in our study, we have also worked
with an aggregate network that includes synapses as well as gap junctions.
Throughout this paper, we have reported results for this weighted combined
network, unless stated otherwise (see Text S1 for the analysis of the
network of synaptic connections).  The link weights in the combined network
correspond to the total number of synaptic and gap junctional connections
from one neuron to another.  The high value of $Q_M$ and dense
inter-connectivity within modules (Fig.~\ref{fig:modular_connectivity},
a) suggest that the network has a modular organization (Table S1).  We
further validate our results by calculating the modularity of randomized
versions of the network where the degree of each node is kept fixed (see
Methods for the network randomization procedure).  The average modularity
of these randomized networks is considerably lower than that of the
empirical network. We examine the robustness of the modular partitioning 
with respect to the method used for detecting communities by using
another module determination algorithm~\cite{Blondel08}. However, this
produces a lower value for $Q (=0.434)$ compared to the spectral method.
The three modules identified using this alternative method have a
substantial overlap (normalized mutual information $I$ = 0.496) 
with the six modules obtained using the spectral method. It suggests
that the alternative algorithm yields a coarser-grained picture of the 
network communities that are considered in the rest of the paper.

The modules do not have a simple relation with the anatomical layout of the
worm. In particular, they are not a result of a simple division of the
nervous system into groups responsible for receiving sensory input, and
other groups involved in motor output. In
Fig.~\ref{fig:modular_connectivity}~(b), we have analyzed the
composition of the different modules in terms of distinct neuron types
(viz. sensory, motor and inter-neurons). None of the modules are
exclusively composed of a single type of neuron, although motor neurons do
tend to dominate one module.

\subsection{Modules and spatial localization}
To understand why modular structures occur in the neuronal network, we
first consider the relation between the optimal partitions and the spatial
localization of neurons in each module. This can tell us whether
constraints related to the physical nearness between neurons, such as
minimization of the wiring length, dictate the topological organization of
the network. Wiring cost has already been shown to be the decisive factor
governing {\em neuron positions} in the body of C.
elegans~\cite{Chen06,Escudero07}.  Thus, a plausible hypothesis is that, if
most neuronal connections occur within a group of neurons which are
physically adjacent to each other, then the wiring cost will be
significantly decreased.  In terms of connectivity, this will be manifested
as a modular organization of the network, where each module will mostly
comprise neurons in close physical proximity.

\begin{figure}[!ht]
\begin{center}
\includegraphics[width=0.65\linewidth]{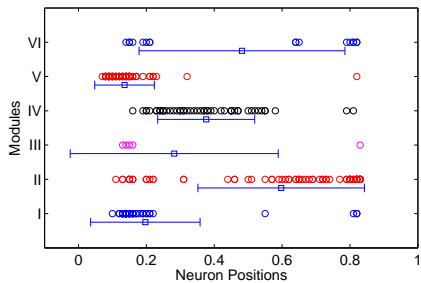}
  \end{center}
	\caption{
	{\bf Neuronal layout of the worm indicating cell body positions of each
	neuron.} The position of neuronal cell bodies along the longitudinal axis
	of the C.  elegans body plan is shown, with the vertical offset and color
	indicating the module to which a neuron belongs.  The mean and standard
	deviation of neuronal positions for each module is also indicated,
	suggesting relative absence of spatial localization in the modules.}
\label{fig:neuron_position}
\end{figure}

Fig.~\ref{fig:neuron_position} indicates the spatial location of the cell
body for each neuron on the nematode body (along the longitudinal axis),
segregated according to their membership in different modules. We see that
a large fraction of the neurons belonging to the same module do indeed have
their cell bodies close to each other. 
A one-way ANOVA test, comparing the positions of the neurons in the 
different modules with the null hypothesis that they are drawn from the same
population, shows that it can be rejected at confidence level of $99.9 \%$.
However, the large  standard
deviations for the distribution of positions of the module components
reveal that none of the modules are spatially localized at any specific
region on the nematode body axis.  
A more detailed multiple comparison procedure carried out for every pair 
of modules shows the
absence of statistically significant spatial separation for the module pairs
(I, III), (I, V), (II, VI), (III, IV), (III, V), (III, VI) and (IV, VI).
The lack of significant distinction
between the modules in terms of their physical location weighs against the
hypothesis of wiring length minimization being the dominant factor
governing the connectivity. 

\begin{figure}[!ht]
\begin{center}
\includegraphics[width=0.85\linewidth]{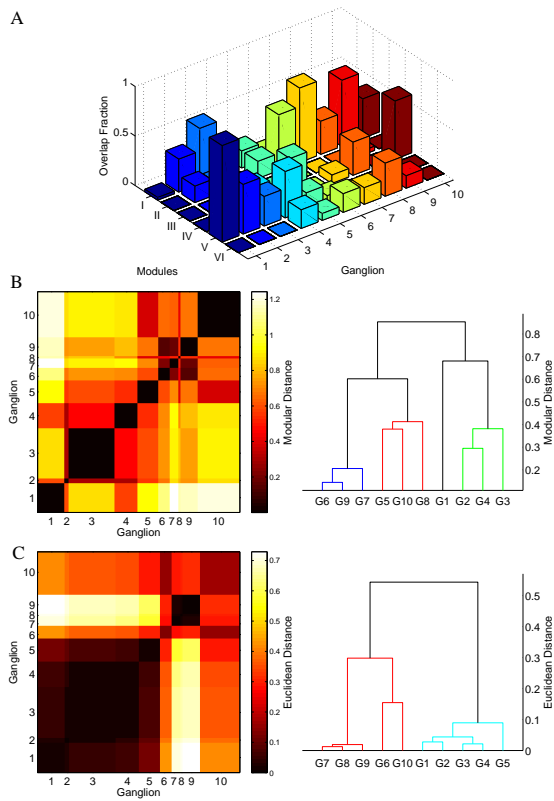}
  \end{center}
	\caption{
	{\bf Modular decomposition of neurons in different ganglia.} (a) Neurons
	belonging to different ganglia are decomposed according to their modular
	membership.  The height of each bar in the histogram corresponds to the
	overlap between the ganglia and the modules, calculated as the fraction
	of neurons that are common to a particular ganglion and a specific
	module.  (b) The matrix representing the average modular distance between
	the different ganglia, as calculated from the modular decomposition
	spectrum of each ganglion.  The corresponding dendrogram indicates the
	closeness between different ganglia in the abstract 6-dimensional
	``modular" space.  (c) The matrix of physical distances between the
	ganglia is shown for comparison with (b), calculated as the average
	distance between neurons belonging to the different ganglia.  The
	corresponding dendrogram indicates the closeness between ganglia
	according to the geographical nearness of their constituent neurons in
	the nematode body. The difference indicates that the ganglia which are
	geographically close may not be neighbors in terms of their modular
	spectra.}
\label{fig:modular_distance}
\end{figure}

The above conclusion is further supported by analysing the connectivity
pattern of the different ganglia of the nematode nervous system.  The nine
anatomically defined ganglia (G1: Anterior, G2: Dorsal, G3: Lateral, G4:
Ventral, G5: Retrovesicular, G6: Posterolateral, G7: Preanal, G8:
Dorsorectal and G9: Lumbar), in addition to the ventral cord (G10), are
defined in terms of physical proximity of their component neurons. Thus, a
lower total connection length between neurons would result in the ganglia
having a relatively higher density of connections between their
constituent neurons. This would imply that the existence of ganglia
imposes a modular structure in the connection topology. To examine how
well the ganglionic arrangement explains the observed modularity of
the neuronal network, we have measured the modularity value $Q_g$ where 
the network communities correspond to the different ganglia.  Although the
non-zero value of $Q_g$ indicates that the connection density between the
neurons in a ganglion is higher than that for the overall network, it is not
as high as the maximum possible value of $Q$ ($= Q_M$, obtained for the
optimal partitioning) as seen from Table~\ref{tab:mod_network}.  To measure
the overlap between the modules obtained by optimal partitioning of the
network and the ganglia, we calculate the normalized mutual information
index, $I$ (see Methods).  In the case of perfect match between the two, $I
= 1$, while, if there is no overlap, $I=0$.  The low values for $I$ given in
Table~\ref{tab:mod_network} suggest that the composition of the different
ganglia is quite distinct from that of the modules. The overlap between
the modules and the ganglia is shown explicitly in
Fig.~\ref{fig:modular_distance}~(a), indicating that most ganglia are
composed of neurons belonging to many different modules.  

This distribution of the neurons of each ganglion into the $m$ different
modules of the optimal partition allows us to define a {\em modular
decomposition} spectrum for the different ganglia. It gives us a metric for
inter-ganglionic distance in a $m$-dimensional ``modular'' space.  Thus, in
this abstract space, two ganglia are close to each other if they have
similar spectral profiles. Their distance in this ``modular" space
(Fig.~\ref{fig:modular_distance},~b) are then compared to their
physical distance, as measured in terms of the average separation between
the cell bodies of all pairs of neurons $i$ and $j$, belonging to different
ganglia (Fig.~\ref{fig:modular_distance},~c). The comparison of the
two distance matrices shows that there are indeed certain similarities
between these two different concepts of closeness between the ganglia.  For
example, the five ganglia located in the head (G1-G5) cluster together, as
do the three located towards the tail (G7-G9). This similarity
can be quantified by computing the correlation between these two distance
measures (i.e., Euclidean and modular), $r = 0.564$ ($p < 0.0001$). 
Our observation is in
accord with  previous reports which use the notion of wiring cost for
explaining (to a certain extent) the observed relative positions of the
ganglia~\cite{Cherniak94}. However, when we consider the corresponding
dendrograms that indicate the relative proximity of the different ganglia
in physical space and in ``modular'' space, we observe significant
differences between the two. Ganglia which are close to each other in
physical space may not be neighbors in terms of their modular spectra. For
instance, G5 which is located in the head, is closer in ``modular" space to
the ganglion G8 located in the tail. On computing the correlation coefficient
between the two trees by considering the distances between every pair 
of ganglia (measured as the path length between the pair in the dendrogram),
we 
obtain $r = 0.516$ ($p<0.0003$). This value is substantially lower than 1,
the value expected had the two dendrograms been identical.  
It reiterates our previous
conclusion that wiring cost minimization, which is related to the physical
distance between neurons, is not a dominant factor governing the
organization of C. elegans somatic nervous system.

\subsection{Modules and cell lineage}
As developmental processes are believed to play a critical role in
determining the structure of the nervous system, we also consider the
alternative hypothesis that the structural modules reflect a clustering of
neurons that are related in terms of their lineage. Lineage of a cell is
the pattern of successive cellular divisions that occur during its
development.  This is invariant in C. elegans, allowing one to trace the
individual developmental history of each cell in order to identify the
cell-autonomous mechanisms and inter-cellular interactions that define its
fate~\cite{CelegansII}.  
We investigate whether a relation exists between the modular
structure and the sequence of cell divisions that occur during development,
by measuring the average relatedness between neurons occurring in the same
module and comparing with that for neurons occurring in different modules.

\begin{figure}[!ht]
\begin{center}
\includegraphics[width=0.65\linewidth]{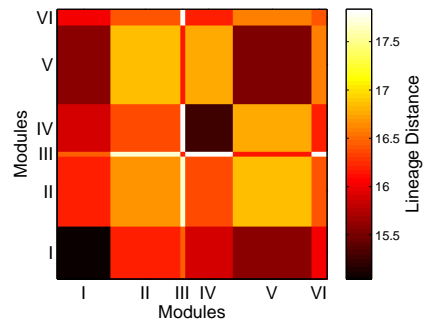}
  \end{center}
	\caption{ 
	{\bf Lineage distance between modules.} The matrix representing the
	average lineage distance between neurons occurring within the same module
	and those belonging to different modules.  The figure indicates that
	neurons occurring in the same module have only a slightly lower lineage
	distance as compared to that between neurons occurring in different
	modules.} 
\label{fig:lineage}
\end{figure}

Fig.~\ref{fig:lineage} indicates that it is difficult to distinguish
the modules in terms of the lineage of the neurons comprising them.
Indeed, even coarse distinctions such as AB and non-AB lineage neurons are
not apparent from the modular division.  
A one-way ANOVA test, with the null hypothesis that the average lineage 
distance between neurons within the same module and that between neurons
belonging to different modules are obtained from the same distribution,
shows that it cannot be rejected at confidence level of $99 \%$.
We next analyse each pair of modules using a multiple comparison procedure 
for 
the intra-modular and inter-modular lineage distances of their constituent 
neurons. This reveals that the neurons belonging to modules III and VI 
have within-module 
lineage distance distribution that is statistically indistinguishable 
from the distribution of their lineage distance with neurons belonging 
to any of the other modules. Thus, for at least two of the modules, one
cannot segregate them in terms of cell lineage.
The detailed view of the
relatedness between each pair of neurons shown in Fig.~S1 indicates that,
while in each module there are subgroups of closely related neurons,
different subgroups within the same module may be very far from each other
in the lineage tree.  Conversely, neurons occurring in different modules
can have small distance between each other in terms of lineage.  
This observation is supported by the low correlation $r = 0.076$
($p < 0.0001$) between the physical and lineage distances of neurons.
The fact
that C.  elegans neurons are largely non-clonally derived from many
different parent cells \cite{Hobert05} may partly explain this lack of
correlation between lineage and modules.  The above results indicate that
developmental constraints arising from common ancestry are not exclusively
responsible for the observed connection structure of the C. elegans
neuronal network. 

\subsection{Optimizing between wiring cost and communication efficiency}
In the previous section, we have shown that neither wiring cost
minimization nor lineage considerations can by themselves determine the
connection topology of the network. In order to ascertain the possible
nature of the additional constraints that gives rise to the observed
mesoscopic structure, we now investigate global properties of the neuronal
network.  A possible governing factor for network organization is that,
rather than decreasing the total wiring length or the average physical
distance between connected neurons, the network minimizes the path length
for information transfer. This can be measured by the number of links that
must be traversed to go from one neuron to another using the shortest
route~\cite{Kaiser06}.  We consider this possibility by measuring the
communication efficiency of the network, using the harmonic average path
length between all pairs of neurons (see Methods).

\begin{figure}[!ht]
\begin{center}
\includegraphics[width=0.85\linewidth]{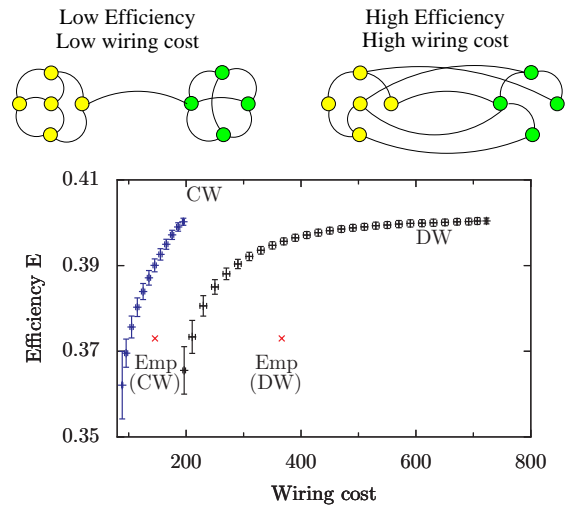}
  \end{center}
	\caption{{
	\bf Trade-off between wiring cost and communication efficiency in the
	network.} The variation of communication efficiency, $E$, as a function
	of the wiring cost, defined using either the ``dedicated-wire" model (DW)
	or the ``common-wire" model (CW), in the ensemble of random networks with
	degree sequence identical to the C. elegans neuronal network. The trend
	indicates a trade-off between increasing communication efficiency and
	decreasing wiring cost. The corresponding values for the empirical
	network are indicated by crosses for both DW and CW. The schematic
	figures shown above the main panel indicate the type of networks obtained
	in the limiting cases when only one of the two constraints are satisfied.
	In both curves, error bars indicate the standard deviations calculated
	for $10^3$ random realizations. We observe that the empirical network is
	suboptimal in terms of wiring cost and communication efficiency,
	suggesting the presence of other constraints governing the network
	organization.}
\label{fig:cost_eff}
\end{figure}

It is evident that increasing the efficiency requires topological long-range
connections, which however increases the wiring cost of the network
(Fig.~\ref{fig:cost_eff} top).  Therefore, it is natural to expect that the
system would try to optimize between these two constraints.  Thus, we
compare the performance of the network as a rapid signal propagation system
against the resource cost for the required number of connections.  This
cost is measured as the Euclidean length between the cell bodies of all
connected pairs of neurons, corresponding to the ``dedicated-wire" model of
Ref.~\cite{Chen06}. It has been shown that the positions of the subset of
sensory and motor neurons directly connected to sensory organs and muscles,
respectively, can be determined quite accurately by minimizing their total
wiring cost~\cite{Escudero07}. As our focus is on the connection structure
of the neuronal network, we keep the neuron positions invariant.  By
randomizing the network, keeping the degree of each node unchanged, we can
construct a system with a specific wiring cost. We then measure its
communication efficiency. Fig.~\ref{fig:cost_eff} reproduces the expected
result that, decreasing the wiring cost of the network causes a decline in
its performance in terms of its ability to propagate signals rapidly.
However, it is surprising that the empirical network has a wiring cost much
higher than that of the corresponding randomized network having the same
communication efficiency. To see whether this could be an artifact of the
measure used to calculate wiring cost, we have considered an alternative
method for quantifying it.

Most of the neurons in C. elegans have at most one or two extended
processes, on which all the synapses and gap junctions with other neurons
are made.  Thus the ``dedicated-wire" definition of wiring cost that sums
Euclidean distances between every connected neuronal pair may be a gross
over-estimate of the actual usage of resources used in wiring. Instead, we
can use a ``common-wire" model to define the wiring cost for connecting to
a specific neuron. This is measured by the Euclidean distance between the
neuron's cell body and that of the farthest neuron (along the longitudinal
axis) it is connected to. The simple one-dimensional simplification of the
C.  elegans body that we have assumed here ignores distance along the
transverse plane. Thus, this measure is actually an under-estimate of the
actual wiring cost, and should provide an insightful comparison with the
above measure obtained from the ``dedicated-wire" model.
Fig.~\ref{fig:cost_eff}~(inset) shows that wiring cost increases with
communication efficiency for the randomized networks, which is
qualitatively similar to the relation obtained using the preceding
definition for wiring cost.  In this case also, we find that the empirical
C. elegans network has a much lower efficiency in comparison with an
equivalent randomized network having the same wiring cost. Thus, as this
observation is independent of these two definitions of wiring cost, it suggests
the presence of other constraints that force the neuronal network to have a
higher wiring cost or lower efficiency than we would have expected. These
constraints are possibly related to information processing, which is the
principal function of the nervous system.

\subsection{Information processing
is a distinctive feature of the C. elegans neuronal network}
To explore further the possibility that the additional constraints
governing the topological structure of C. elegans nervous system may be
related to information processing, we investigate how this functional
requirement could be responsible for differentiating the system from other
complex networks for which communication efficiency is of paramount
importance. Rapid communication of information between different
neurons is certainly an important performance criterion. However, the
neuronal network has properties quite distinct from that of (say) the
Internet or the airline transportation network, which are systems designed
for maximum transportation  efficiency of signals or physical resources.

\begin{figure}[!ht]
\begin{center}
\includegraphics[width=0.95\linewidth]{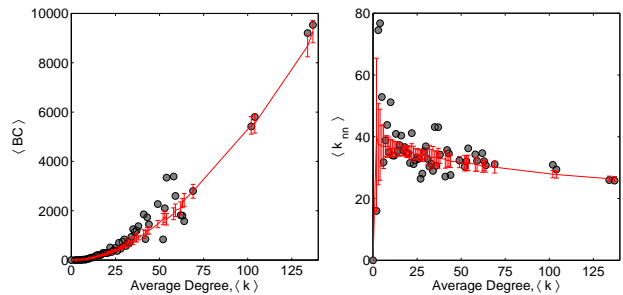}
  \end{center}
	\caption{
	{\bf Betweenness centrality and the average nearest neighbor degree as a
	function of the total degree of network.} (a) The average betweenness
	centrality, $\langle BC \rangle$, and (b) the average nearest neighbor
	degree, $\langle k_{nn} \rangle$ of each node as a function of its total
	degree, $\langle k \rangle = \langle k_{\rm in} + k_{\rm out} \rangle$.
	Betweenness centrality is a measure of how frequently a particular node
	is used when a signal is being sent between any pair of nodes in the
	network using the shortest path. In case of the Internet, BC of nodes
	increases with its degree which is sought to be linked with its
	information transport property.  In C. elegans, although BC increases
	with degree, this increase is not significant when compared to the
	randomized version of the network. In the case of the relation between
	the average connectivity of nearest neighbors of a node with its total
	degree $k$, we note that for both the Internet and protein interaction
	network, $k_{nn}$ decreases with $k$ as a power law.  This means that low
	connectivity nodes have high degree nodes as their neighbors and
	vice-versa. However, in the case of C. elegans, this relation is not very
	apparent and insignificant in comparison with the randomized version of
	the network.  In both figures, error bars indicate the standard
	deviations calculated for $10^3$ random realizations. These results
	suggest that the C.elegans network forms a class distinct from the class
	of networks optimized only for signal propagation.}
\label{fig:deg_vs_bc}
\end{figure}

For this purpose, we look at the overall network design by decomposing the
system into (i) a strongly connected component (SCC), within which it is
possible to visit any node from any other node using directed links, (ii)
an inward component (IN) and (iii) an outward component (OUT), consisting
of nodes from which the SCC can be visited or which can be visited from the
SCC, respectively, but not vice versa. In addition, there can be
disconnected components, i.e., nodes which cannot be visited from SCC nor
can any visits be made to SCC from there (Fig.~\ref{fig:celegans}). A
comparison of the C.  elegans neuronal network with a similar decomposition
of the WWW~\cite{Broder00} reveals that, while in the latter the different
components are approximately of equal size (WWW: SCC $\sim$ 56 million pages, 
while IN, OUT consist of $\sim$ 43 million pages each), the SCC of the 
nervous system comprises almost the entire network (SCC has 274 neurons, IN
has 4 neurons and OUT has 1 neuron).  
Thus, any node can, in principle,
affect any other node in the nervous system, suggesting the importance of
feedback control for information processing.

Next, we consider the relation between two fundamental properties of the
network: the degree of nodes and their {\em betweenness centrality} (BC),
which characterizes the importance of a node in information propagation
over the network (see Methods).  We observe that for both the C. elegans
neuronal network and its randomized versions, the degree of a node and its
BC are strongly correlated, i.e., highly connected nodes are also the most
central (Fig.~\ref{fig:deg_vs_bc},~left).  This is similar to what has been
observed in the Internet~\cite{Vazquez02}, where the highest degree nodes
are also those with the highest betweenness~\cite{Goh03}, but in sharp
contrast to the airport transportation network, where non-hub nodes (low
degree) may have very large BC~\cite{Guimera05a}.

However, the C. elegans neuronal network differs from 
other networks whose  primary function is to allow signal propagation between 
nodes (viz., the Internet and the protein interaction network (PIN)),
in  terms of the variation of the degree
of a node with the average degree of its neighboring nodes, $\langle k_{nn}
\rangle$. While in the Internet and PIN, $\langle k_{nn} \rangle$ decays as
a power law with node degree, in the neuronal network, this dependence is
very weak (Fig~\ref{fig:deg_vs_bc}, right), especially when contrasted 
with the randomized ensemble. This implies that the C. elegans nervous system
does not have multiple star-like subnetworks as seen in the
Internet and PIN~\cite{Guimera07}.
Further, it is different from the
airline transportation network, where the high degree nodes are closely
connected among themselves showing an assortative behavior~\cite{Barrat04}.
In fact, computation of the assortativity coefficient $r = - 0.093$
indicates that the network is disassortative (as previously stated
in Refs.~\cite{Newman02a,Newman03c}), although comparison with that
of the degree-conserved randomized ensemble ($r_{rand} = - 0.078 \pm 0.011$)
indicates that this is predominantly a result of the degree sequence.
Another important distinguishing characteristic of the C. elegans network
is that the distributions of link weight and degree do not appear to be
scale-free, or even having a long tail (Fig.~S2), unlike systems such as
the Internet and the airline transportation
network~\cite{Satorras01a,Barrat04}. Thus, our study shows that there are
additional constraints governing the nervous system connection topology in
C. elegans, which are unrelated to wiring cost, lineage or communication
efficiency. As the principal function of the system is to process
information, the above results suggest it is this functional requirement
that provides the additional constraints leading to the observed
organization of the nematode neuronal network. 

\subsection{Modules and functional circuits}
In order to understand the nature of functional considerations that may
govern the network organization, we focus on the overlap of several
previously identified functional circuits of C. elegans with the
structurally identified modules.  Functional circuits are a subset of
neurons which are believed to play a vital role in performing a function,
and are distinguished by observing abnormal behavior of the organism when
they are individually removed (e.g., by laser ablation). In biological
systems, it has been observed that members of structurally defined
communities are often functionally related (e.g., in the intra-cellular
protein interaction network~\cite{Guimera05} and the network of cortical
areas in the brain~\cite{Hilgetag00,Bassett06}).  Here, we investigate the
possibility of a similar correlation between the anatomical modules of C.
elegans and its functional circuits.  We consider the functional circuits
for (F1) mechanosensation~\cite{Chalfie85,Wicks96,Sawin96}, (F2) egg
laying~\cite{Waggoner98,Bany03}, (F3) thermotaxis~\cite{Mori95}, (F4)
chemosensation~\cite{Troemel97}, (F5)
feeding~\cite{Gray05,Chalfie85,White86}, (F6)
exploration~\cite{Gray05,Chalfie85,White86} and (F7) tap
withdrawal~\cite{Wicks95,Wicks96} (Table S2).

\begin{figure}[!ht]
\begin{center}
\includegraphics[width=0.95\linewidth]{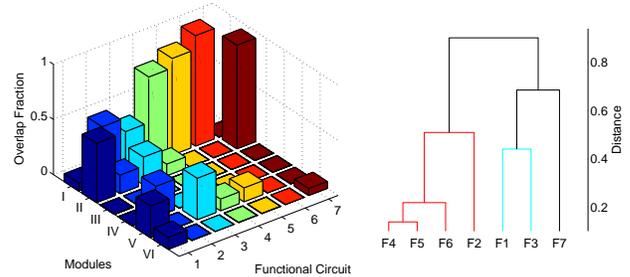}
  \end{center}
	\caption{
	{\bf Modular decomposition of neurons in different functional circuits.}
	Neurons belonging to different functional circuits are decomposed
	according to their modular membership.  The height of each bar in the
	histogram corresponds to the overlap between the modules and functional
	circuits (F1) mechanosensation, (F2) egg laying, (F3) thermotaxis, (F4)
	chemosensation, (F5) feeding, (F6) exploration and (F7) tap withdrawal.
	The overlap is measured in terms of the fraction of neurons common to a
	particular functional circuit and a specific module.  The corresponding
	dendrogram represents the closeness between different functional circuits
	in the abstract 6-dimensional ``modular" space.}
\label{fig:overlap_fn}
\end{figure}

Fig.~\ref{fig:overlap_fn} shows the modular decomposition of each
functional circuit, indicating their overlap with the modules. The
corresponding dendrogram clusters the circuits in terms of the similarity
in their modular spectra. We note that the circuits for chemosensation,
feeding and exploration are clustered together.  This is consistent with
the fact that most of the neurons belonging to the feeding and exploration
circuits are involved in chemosensation.  A surprising observation is that
although F6 is a subset of F4, it is actually closer to F5 in ``modular"
space (distance = 0.18) than to F4 (distance = 0.26), despite the feeding
and exploration circuits not having any neuron in common.  This close relation
between F5 and F6 in modular space is suggestive of a relation between
modularity and functionality, as it is known from experiments that there
is a strong connection
between their corresponding functions. The feeding behavior of
C.elegans is known to be regulated in a context-dependent manner by its
chemical milieu.  It integrates external signals~\cite{Daniels08}, such as
the availability of food, and nutritional status of the animal, to direct
an appropriate response~\cite{Franks06}.  An example is the avoidance of
high CO$_2$ concentrations by satiated animals~\cite{Bretscher08}.
Further, the mode of locomotion of the organism is also determined by the
quality of food~\cite{Shtonda06}.  Another important observation made from
the modular decomposition is the proximity of the functional circuit F2 to
the group (F4,F5,F6). This is significant in light of experimental
observations that presence of food detected through chemosensory neurons
modulates the egg-laying rate in C.  elegans~\cite{Sawin96,Sawin00}.  The
above results indicate that the relation between the functional circuits,
which are essential for the survival of the organism, are reflected in the
modular organization of the nematode nervous system.

\subsection{Functional roles of different neurons}
Having looked at the functional circuits and the relations between them in
the previous section, we now investigate the importance of individual
neurons in terms of their connectivity. This is revealed by a comparison
between the localization of their connections within their own community
and their global connectivity profile over the entire network.  In order to
do this, we focus on (i) the degree of a node within its module, $z$, that
indicates the number of connections a node has to other members of its
module, and (ii) its participation coefficient, $P$, which measures how
dispersed the connections of a node are among the different
modules~\cite{Guimera07}.

A node having low within-module degree is called a non-hub ($z<0.7$) which
can be further classified according to their fraction of connections with
other modules. Following Ref.~\cite{Guimera07}, these are classified as
(R1) ultra-peripheral nodes ($P \leq 0.05$), having connections only
within their module, (R2) peripheral nodes ($0.05 < P \leq 0.62$), which
have a majority of their links within their module, (R3) satellite
connectors ($0.62 < P \leq 0.8$), with many links connecting nodes outside
their modules, and (R4) kinless nodes ($P > 0.8$), which form links
uniformly across the network. Hubs, i.e., nodes having relatively large
number of connections within their module ($z \geq 0.7$), are also divided
according to their participation coefficient into (R5) provincial hubs ($P
\leq 0.3$), with most connections within their module, (R6) connector hubs
($0.3 < P \leq 0.75$), with a significant fraction of links distributed
among many modules, and (R7) global hubs ($P > 0.75$), which connect
homogeneously to all modules. This classification allows us to distinguish
nodes according to their different roles as brought out by their
intra-modular and inter-modular connectivity patterns (Table S3). 

\begin{figure}[!ht]
\begin{center}
\includegraphics[width=0.95\linewidth]{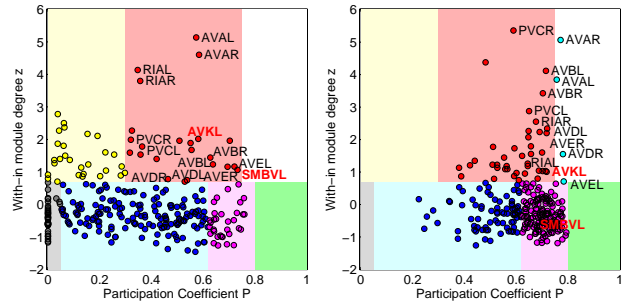}
  \end{center}
	\caption{
	{\bf The role of individual neurons according to their intra- and
	inter-modular connectivity.} (a) The within module degree $z$-score of
	each neuron in the empirical neuronal network is shown against the
	corresponding participation coefficient $P$.  The within module degree
	measures the connectivity of a node to other nodes within its own module,
	while the participation coefficient measures its connectivity with
	neurons in the entire network.  (b) The corresponding result for a
	randomized version of the C. elegans network where the degree of each
	neuron is kept unchanged is also shown.  Neurons belonging to the
	different regions in the $P-z$ space are categorised as: (gray)
	``ultraperipheral nodes,'' i.e., nodes with all their links within their
	module, (blue) ``peripheral nodes,'' i.e., nodes with most links within
	their module, (pink) ``nonhub connector nodes,'' i.e., nodes with many
	links to other modules, (green) ``nonhub kinless nodes,'' i.e., nodes
	with links homogeneously distributed among all modules, (yellow)
	``provincial hubs,'' i.e., hub nodes with the vast majority of links
	within their module, (red) ``connector hubs,'' i.e., hubs with many links
	to most of the other modules, and (white) ``global hubs,'' i.e., hubs
	with links homogeneously distributed among all modules. The neurons
	occurring as connector hubs are identified in the figure. Most of these
	neurons occur in different functional circuits indicating the close
	relation between functional importance and connectivity pattern of
	individual neurons. In addition, the neurons AVKL and SMBVL which are
  predicted to be functionally important are separately marked.}
\label{fig:global}
\end{figure}

We will now use the above methodology on the C. elegans network in order to
identify neurons that play a vital role in coordinating activity through
sharing information (either locally within their community or globally over
the entire network).  Fig.~\ref{fig:global} shows the comparison between
the empirical network and a corresponding randomized network (obtained by
keeping the degree of each node fixed). 
Results for a randomized ensemble, comparing the number of neurons
in each role against that for the empirical network, are given in Table S4.
We immediately notice that the
randomized networks have relatively very few nodes having the roles R1 and R5,
indicating that the modular nature of the original network has been lost.
In fact, in the randomized system, most nodes have higher participation
coefficient, with a large majority being satellite connectors (R3). More
interesting is the fact that, the empirical neural network does not possess
any neuron having the global roles played by R4 and R7, whereas these
regions may be populated in randomized networks. This implies that modular
identity in the C.  elegans neuronal network is very pronounced.

It is possible that, neurons having the role of provincial hubs may be
involved in local coordination of neural activity, while, the connector
hubs may be responsible for integration of local excitations to produce a
coherent response of the entire system. This hypothesis is supported by
noting that all command interneurons (of the class AVA, AVB, AVD, AVE,
PVC), which control forward and backward locomotion of the worm by
regulating motor output, play the role of connector hubs.  In fact, out of
the 23 neurons in the class R6, 20 are known to belong to different
functional circuits.  Among the rest, although DVA does not belong to any of
the known circuits, it has recently been identified as being involved in
mechanosensory response and in its absence, the frequency and magnitude of
the tap-induced reversal, as well as the acceleration magnitude, is
diminished~\cite{Wicks95}.  The two remaining neurons, AVKL and SMBVL, have
not been implicated so far in any known functional circuit. However, their
occurrence in this class suggests that they may be important for some, as
yet unknown, function. This is a potentially interesting prediction that
may be verified in the laboratory.  

The significance of these results is underlined by a comparison with the
randomized network.  For instance, in the random realization shown in
Fig.~\ref{fig:global}~(b), of the 49 neurons playing the role of connector
or global hubs, less than half (viz., 23) actually belong to any of the
known functional circuits.  The appearance of most of the command
interneurons in the high-$z$ region of both the empirical and randomized
networks indicates that their high overall degree is responsible for their
observed role of ``connecting hubs". 

We now turn to the 28 neurons which play the role of provincial hubs.  Half
of all the inhibitory D-class motorneurons (viz., DD1-DD3 and VD1-VD6) are
found to belong to this class. This is significant as these neurons have
already been implicated in the ability of the worm to initiate backward
motion. While they also contribute to forward locomotion, previous
experiments have shown that they are not essential~\cite{McIntire93}.  This
fits with our hypothesis that, R5 neurons are important for local
coordination but may not be crucial for the global integration of activity.
A pair of excitatory B-class motorneurons that sustain coordinated forward
locomotion in the worm also appear as provincial hubs. Of the remaining R5
neurons, 9 have been previously identified as belonging to various
functional circuits. It will be interesting to verify the functional
relevance of the remaining 8 neurons (OLLL/R, RMDVL/R, SMDVR, RIH, RMDDL/R)
in the laboratory.  Thus, overall, we find a very good correlation between
the connectivity pattern and the functional importance of different
neurons.

Analysis of neurons having different roles in terms of their membership in
the different ganglia (Fig. S3) indicates that the lateral ganglion
provides the majority of neurons acting as connector hubs (R6). This is
consistent with an earlier study where this ganglion was found to be the
principal highway for information flowing between neurons responsible for
receiving sensory stimuli and those involved in motor
response~\cite{Chatterjee07}.  To check the significance of the above
result, we observe the membership of the set of connector hubs in the
corresponding randomized network (Fig.  S3).  While this also shows many
neurons from the lateral ganglion, unlike in the empirical network it has
significant representation from other ganglia too (e.g., the retrovesicular
ganglion). 

We have also carried out an analysis of the frequency of links between
neurons having different roles, relative to the randomized network (Fig.
S4). This allows us to compare the C. elegans neuronal network with other
networks involved in (a) transportation and (b) information propagation. It
has been shown that networks of class (b) shows significant
under-representation of links between R1-R1, R5-R6 and R6-R6, whereas
networks of class (a) exhibit over-representation of all
three~\cite{Guimera07}. These patterns have been related to the occurrence
of stringy periphery in class (a) and multi-star structures in class
(b)~\cite{Guimera07}. In the case of C. elegans, R1-R1 does seem to be
over-represented. However, R6-R6 shows very little over-representation,
while both R5-R6 and R6-R5 show slight under-representation. This
difference in the role-to-role connectivity pattern for the nematode
nervous system with the networks in the above-mentioned two classes
suggests that its
structure is not exclusively characterized by either a stringy periphery or
multiple stars. This assumes significance in light of recent work
distinguishing information (or signalling) networks, such as the Internet
and protein interactome, on the one hand, and transportation networks, such
as metabolic and airport networks, on the other, into two
classes~\cite{Guimera07}. Our results suggest that neuronal networks which
have to {\em process} information, in addition to transferring signals, may
constitute a different category from either of the above classes.

\section{Discussion}
In this paper, we have carried out a detailed analysis of the mesoscopic
structure in the connection topology of the C. elegans neuronal network.
Inferring the organizing principles underlying the network may give us an
understanding of the way in which an organism makes sense of the external
world. We have focused primarily on the existence of modules, i.e.,
groups of neurons having higher connection density among themselves than
with neurons in other groups. The presence of such mesoscopic organization
naturally prompts us to ask the reasons behind the evolution of these
features in the network.

In lower invertebrates like nematodes, the genome is the dominant factor
which governs the development of the organism, including its nervous
system. The neuronal network structure is formed early in the life-cycle of
the organism, when most of the cells and their connections are configured
permanently.  Although external cues may play a role, the relative absence
of individual variations in the network organization makes C. elegans an
ideal system for studying how the system has evolved to optimize for
various constraints, such as minimizing resource use and maximizing
performance.

There have been recent attempts at explaining neuronal position and
structural layout of the network by using static constraints, such as
wiring economy and communication path minimization.  Although we find that
membership of neurons in specific modules are correlated with their
physical nearness, the empirical network is sub-optimal in terms of both
the above-mentioned constraints.  By comparing the system with other
complex networks that have been either designed or have evolved for rapid
transportation while being subject to wiring economy, we find that the C.
elegans nervous system stands apart as a distinct class. This suggests that
the principal function of neuronal networks, viz., the processing of
information, distinguishes it from the other networks considered, and plays
a vital role in governing its arrangement.  Considering the importance of
this constraint in ensuring the survival of an organism, it is natural that
this should be key to the organizing principles underlying the design of
the network.  The intimate relation between function and structure of the
nervous system is further brought out by our use of structural analysis to
distinguish neurons that are critical for the survival of the organism. In
addition to identifying neurons that have been already empirically
implicated in different functions (which serve as a verification of our
method), we also predict several neurons which can be potentially crucial
for certain, as yet unidentified, functions.

Biological systems are distinguished by the occurrence of discrete entities
with a distinct function, which are often termed as {\em functional}
modules~\cite{Hartwell99}.  For several networks that occur in the
biological context (such as that of protein interactions), the components
of {\em structural} modules are seen to be functionally related.  This
suggests that modules provide a framework for relating mesoscopic patterns
in the connection topology to subsystems responsible for specific
functions.  Although there is no unique correspondence between the
structural modules of C. elegans and the known functional
circuits~\cite{Arenas08}, we use the overlap of the circuits with the
modular membership of their constituent neurons to discover correlations
between them.  Our results reveal non-trivial association between circuits
whose corresponding functions are closely connected, even when they do not
share any common neurons. As such relations  could not have been revealed
by a micro-scale study which focuses on individual neurons and their
connections, this result highlights one of the significant advantages of
investigating the network at the mesoscopic level.

When we compare the nervous system of C. elegans with the brains of higher
organisms, we observe the modular organization of the latter to be more
prominent~\cite{Muller-Linow08}. For example, the network of cortical areas
in the cat and macaque brains exhibit distinct
modules~\cite{Zhou06,Pan09a}, with each module being identified with
specific functions~\cite{Hilgetag00,Bassett06}.  A possible reason for the
relatively weak modular structure in the nematode could be due to the
existence of extended processes for the neurons of C. elegans. Many of
these span almost the entire body length, an effect that is enhanced by the
approximately linear nature of the nematode body plan.  As a result,
connections are not constrained by the physical distance between soma of
the neurons, as is mostly the case in mammalian brains. It is apparent that
such constraints on the geographical distance spanned by links between
nodes (viz., cost of wiring length) can give rise to clustering of
connections among physically adjacent elements.  In addition, the small
nervous system of C. elegans, comprising only 302 neurons, lacks
redundancy. Therefore, individual neurons may often have to perform a set
of tasks which in higher organisms are performed by several different
neurons.  Thus, functional modularity is less prominent in the nematode, as
some neurons belong to multiple behavioral circuits.

Another principal distinction between the C. elegans nervous system and the
brains of higher organisms such as human beings, is the relative high
connectivity in the former (the connection density being $C \sim 0.1$).  By
contrast, the connectance for human brain is around
$10^{-6}$~\cite{Shepherd98}, which leads us to the question of how
communication efficiency can remain high in such a sparsely connected
network. It is possible that the more intricate hierarchical and modular
structures seen in the brains of higher organisms is a response to the
above problem. The fact that the rate at which the number of neurons $N$
increase across species, is not matched by a corresponding increase in the
number of links (which increases slower than $N^2$) implies the existence
of constraints on the latter, which is a resource cost in addition to the
earlier mentioned cost of wiring length.
Note that, in the present work we have focused on a single level of modular
decomposition of the nematode neuronal network. It is possible that the
system may have multiple levels of hierarchically arranged
inter-nested modules~\cite{Ferrarini08,Sales-Pardo07}. Investigating
the existence of such organization in the C. elegans nervous system is a 
potential topic for future exploration.

Networks provide the scaffolding for the computational architectures that
mediate cognitive functions. The pattern of connections of a neuron (or a
neuronal cluster) defines its functionality not just locally but also as an
integrated part of the nervous system. This is because neurocognitive
networks across the evolutionary tree consist of interconnected clusters of
neurons that are simultaneously activated for generating a single or a series 
of cognitive outputs. However, while some of the neurons (or clusters of 
neurons)  are essential for the relevant outcome, others are ancillary. Thus,
 they work in a collaborative mode but are not interchangeable, each displaying
relative specializations for separate behavioral components. Since the most
prominent neuronal pathways are those that interconnect components of
functional circuits, our analysis validates the intrinsic connection between
network structure and the functions of the nervous system. The concept that
behavioral consequences of damaging a region of the brain will reflect the
disruption of the underlying network architecture may be intuitive but is
particularly important when one considers brain dysfunction through
neurodegeneration~\cite{Horwitz95}, where atrophy is seen to propagate 
preferentially through networks of functionally related 
neurons~\cite{Seeley09}. While the idea that
damage in one part of the brain can affect other areas connected to that
region is not new, our work on the mesoscopic network organization may be
extended to look at disease progression through the network, beyond the
current focus on the pathology within individual cells.

% You may title this section "Methods" or "Models". 
% "Models" is not a valid title for PLoS ONE authors. However, PLoS ONE
% authors may use "Analysis" 
\section{Materials and Methods}
{\bf Connectivity Data:}
We have used information about the network connectivity, positions of
neurons and the lineage distance between cells for the C. elegans nervous
system, from the database published in Ref.~\cite{Chen06} and available
from the online Atlas of C. elegans anatomy (www.wormatlas.org).  This is
an updated and revised version of the wiring data originally published in
Ref.~\cite{White86}.  The connectivity between neurons and the positions of
the neuron cell bodies along the longitudinal axis of the worm is
reconstructed based on serial section electromicrography.  Note that, the
new database adds or updates about 3000 connections to the previous
version.  The lineage data indicates the relatedness between every pair of
neurons in terms of distances in the embryonic and post-embryonic lineage
trees.  This is measured by identifying the last common progenitor cells of
the two neurons, and then counting the number of cell divisions from this
common ancestor. Each cell division adds a single unit to the total
lineage distance, with the initial division from the common 
progenitor counted only
once.

{\bf Modularity (Q):}
To decompose a given network into modules (where a module or community is
defined as a subnetwork having a higher density of connections relative to
the entire network) we use a method introduced in Ref.~\cite{Newman04}. We
compute a quantitative measure of modularity, $Q$, for a given partitioning
of the network into several communities, 
\begin{equation}
	Q\equiv \frac{1}{2L}\sum_{i,j}\left[A_{ij} - \frac{k_i
	k_j}{2L}\right]\,\delta_{c_i c_j}.
\label{eqn:modularity}
\end{equation}
Here, $\bf A$ is the adjacency matrix ($A_{ij}$ is 1, if neurons $i$, $j$
are connected, and $0$, otherwise). The degree of each node $i$ is given by
$k_i = \sum_{j=1} A_{ij}$. $L$ is the total number of links in the
network, $\delta$ is the Kronecker delta ($\delta_{ij} = 1$, if $i=j$, and
$0$, otherwise), and $c_i$ is the label of the community to which vertex
$i$ is assigned. In the case of directed and weighted network, the above
measure can be generalized as
\begin{equation}
	Q^{W}\equiv \frac{1}{L^W}\sum_{i,j}\left[W_{ij}-
	\frac{s_{i}^{\rm in} s_{j}^{\rm out}}{L^W}\right]\,\delta_{c_i c_j},
	\label{eqn:weimod}
\end{equation}
where, $L^W=\sum_{i,j} W_{ij}$ is the sum of weights of all links in the
network ($W_{ij}$ is the weight of the link from neuron $j$ to neuron $i$),
and the weighted in-degree and out-degree of node $i$ are given by
$s_{i}^{\rm in} = \sum_j W_{ij}$ and $s_{i}^{\rm out} = \sum_j W_{ji}$,
respectively.  

The optimal partitioning of the network is the one which maximizes the
modularity measure $Q$ (or $Q^{W}$).  We obtain this using a generalization
of the spectral method~\cite{Leicht08,Newman06}. We first define a
modularity matrix  $B$,
\begin{equation}
	B_{ij} = W_{ij} - \frac{s_{i}^{\rm in} s_{j}^{\rm out}}{L^W}.
\end{equation}
To split the network into modules, the eigenvectors corresponding to the
largest positive eigenvalue of the symmetric matrix ($\mathbf B + \mathbf
B^{\mathbf T}$) is calculated and the communities are assigned based on the
sign of the elements of the eigenvector. This divides the network into two
parts, which is refined further by exchanging the module membership of each
node in turn if it results in an increase in the modularity.  The process
is then repeated by splitting each of the two divisions into further
subdivisions. This recursive bisection of the network is carried out until
no further increase of $Q$ is possible. 

{\bf Modular spectra:} 
We analyze different neuronal groups, defined in terms of functions,
anatomy (e.g., ganglia), etc., by proposing a decomposition in terms of the
overlap of their constituent neurons with the different modules. Let the
set of all neurons be optimally partitioned into $m$ modules. We
then define an overlap matrix, $O$, where the rows correspond to the different
neuronal groups, and the columns correspond to the different modules. An
element of this overlap matrix, $O_{ij}$, is the number of neurons in 
group $i$ that are from the module $j$. Then, the decomposition of the
$i$-th group in the abstract $m$-dimensional basis space formed by the
modules is $\{ \frac{O_{i1}}{N_i}, \frac{O_{i2}}{N_i}, \ldots,
\frac{O_{im}}{N_i} \}$, where $N_i = \Sigma_{k=1}^m O_{ik}$ 
is the total number
of neurons in the $i$-th group. The distance between two groups $i$ and
$k$ in this ``modular" space is defined as 
\begin{equation}
d_{ij}^{modular} 
= \sqrt{\sum_k \left[ \frac{O_{ik}}{N_i} - \frac{O_{jk}}{N_j} \right]^2}.
\end{equation} 
Thus, this measure can be used as a metric for closeness between
different neuronal groups.

{\bf Decomposition of the network into SCC, IN and OUT components:}
In order to determine the Strongly Connected Component (SCC) of the
network, we first calculate the graph distance matrix containing the shortest
directed path between every pair of nodes in the network.
A finite path length from neuron $i$ to neuron $j$ indicates the
existence of a connected path from one neuron to the other.
By grouping together all nodes which have finite path length with 
all other members of the group, we determine the SCC. In general, one
can use Tarjan's algorithm for SCC determination~\cite{Tarjan72}.
Next, we identify all neurons not belonging to SCC but which can be reached
via a directed path starting from a node in SCC. 
This constitutes the OUT component of the network. Similarly, the group
of neurons which do not belong to SCC but which have finite directed path
length to a member of SCC, constitutes the set of IN neurons.

{\bf Betweenness centrality (BC):}
To measure the importance of a node in facilitating communication across a
network, we consider how frequently a node is used to convey information
from any part of the network to any other part using the shortest available
path. The betweenness centrality of a node $i$ is defined as the fraction
of shortest paths between the pairs of all other nodes in the network that
pass through $i$~\cite{Freeman76}. If the total number of shortest paths
between nodes $j$ and $k$ is $\sigma_{jk}$, of which $\sigma_{jk}(i)$ paths
pass through node $i$, then the betweenness centrality of node $i$ is
\begin{equation}
	C_B(i) = \sum_{j\rightarrow k} \frac{\sigma_{jk}(i)}{\sigma_{jk}}.
\end{equation}

{\bf Communication Efficiency (E):}
To measure the speed of information transfer over the network, one can
define the efficiency $\varepsilon_{ij}$ of communication between vertices
$i$ and $j$ to be inversely proportional to the shortest graph distance
$\varepsilon_{ij}=1/d_{ij} \forall i,j$. Therefore, the efficiency of
communication across the whole network is
\begin{equation}
E({\mathbf G})=\ell^{-1}=\frac{1}{\frac{1}{2}N(N-1)}\sum_{i>j}\frac{1}{d_{ij}}.
\end{equation}
This is the harmonic mean of graph distances between all pairs, which does
not diverge even when the network is disconnected~\cite{Latora01}.  

{\bf Network Randomization:}
An ensemble of randomized versions of the empirical network is constructed
keeping the in-degree and out-degree of each node unchanged.  Each such
network is created by rewiring randomly selected pairs of directed edges,
$i \rightarrow j$ and $k \rightarrow l$, such that, in the randomized
network, the corresponding directed edges are $i \rightarrow l$ and $k
\rightarrow j$. However, if these new links already exist in the empirical
network, this step is aborted and a new pair of edges are chosen in order
to prevent the occurrence of multiple edges~\cite{Maslov02}. The above
procedure is repeated $5 \times 10^6$ times for a single realization of the
randomized network. In order to compare the properties of the empirical
network with its randomized version, an ensemble of $10^3$ realizations is
considered.

{\bf Determining the intra- and inter-modular role of a neuron:}
The role played by each neuron in terms of its connectivity within its own
module and in the entire network is determined according to two
properties~\cite{Guimera05a}:
(i) the relative within module degree, $z$, and (ii) the participation
coefficient, $P$. 

The $z$-score of the within module degree distinguishes nodes that are hubs
of their communities from those that are non-hubs. It is defined as 
\begin{equation}
	z_{i}=\frac{\kappa_{c_i}^{i}-\langle \kappa_{c_i}^{j} \rangle_{j \in
	c_i}}{\sqrt{\langle (\kappa_{c_i}^{j})^2 \rangle_{j \in c_i} - \langle
	\kappa_{c_i}^j \rangle^2_{j\in c_i}}},
	\label{eqn:zscore}
\end{equation}
where $\kappa_c^i$ is the number of links of node $i$ to other nodes in its
community $c$ and $\langle \cdots \rangle_{j\in c}$ are taken over all
nodes in module $c$. The within-community degree $z$-score measures how
well-connected node $i$ is to other nodes in the community.

The nodes are also distinguished based on their connectivity profile over
the entire network, in particular, their connections to nodes in other
communities. Two nodes with same within module degree $z$-score can play
different roles, if one of them has significantly higher inter-modular
connections compared to the other. This is measured by the participation
coefficient $P_i$ of node $i$, defined as
\begin{equation}
	P_{i}=1-\sum_{c=1}^{m}\left( \frac{\kappa_c^i}{k_i} \right)^2,
	\label{eqn:PC}
\end{equation}
where $\kappa_c^i$ is the number of links from node $i$ to other nodes in its
community $c$ and $k_i=\sum_c \kappa_c^i$ is the total degree of node $i$.
Therefore, the participation coefficient of a node is close to 1, if its
links are uniformly distributed among all the communities, and is 0, if all
its links are within its own community.

{\bf Normalized mutual information ($I$):}
To measure the overlap of the membership of nodes in a ganglion with their
membership in a specific module, we use the normalized mutual information
measure~\cite{Kuncheva04,Danon05}. First, we define a overlap matrix, $O$,
where the rows correspond to the different ganglia, and the columns
correspond to the modules obtained by optimal partitioning of the network.
An element of this overlap matrix, $O_{ij}$, is the number of neurons in the
ganglion $i$ that appear in the module $j$. An information-theoretic
measure of similarity between the partitions can then be defined as
\begin{equation}
	I=\frac{-2\sum_{i=1}^{m_A} \sum_{j=1}^{m_B} O_{ij} \log
	(O_{ij}O/O_{i.}O_{.j})}
	{\sum_{i=1}^{m_A}O_{i.}\log(O_{i.}/O) + \sum_{j=1}^{m_B}O_{.j}\log(O_{.j}/O)},
\end{equation}
where $m_A$ and $m_B$ are the numbers of ganglia and modules respectively.
The sum over row $i$ of matrix $O_{ij}$ is denoted by $O_{i.}$ and the sum
over column $j$ is denoted by $O_{.j}$.  If the modules are identical to
the ganglia, then $I$ takes its maximum value of 1. On the other
hand, if the modular partitioning is independent of the division of the
network into ganglia, $I \sim 0$.

% Do NOT remove this, even if you are not including acknowledgments
\section{Acknowledgments}
We would like to thank R. Anishetty, S. Koushika and V. Sreenivasan for
helpful comments and discussion. 

\section{Supporting Information}
{\bf Figure S1.} 
Matrix representing the relatedness of neurons in the somatic nervous
system of C. elegans as measured in terms of their lineage distance. The
neurons are arranged according to the modules they belong to. The module
boundaries are indicated in the figure.  Within each module, neurons that
are close in terms of lineage are placed in adjacent positions. The figure
shows that closely related neurons may occur in different modules, while
those in the same module may be far apart in terms of lineage distance.
This indicates that there is no simple relation between relatedness of
neurons in terms of lineage and their modular membership.

{\bf Figure S2.}
Cumulative distributions of the strength and (inset) degree for the (a)
gap-junctional, (b) synaptic and (c) combined networks. The gap-junctional
network is undirected and the strength of a node is defined as $s_i=\sum_j
W_{ij}$, where $W_{ij}$ is the number of gap junctions between neurons $i$
and $j$.  On the other hand, the synaptic and combined networks are
directed and the inward- and outward-strength of a node are defined as
$s_i^{\rm in}=\sum_j W_{ji}$, and $s_i^{\rm out}=\sum_j W_{ij}$,
respectively.  For directed networks, $W_{ij}$ represents the number of
connections from neuron $j$ to $i$.  The figures indicate that scale-free
behavior of the distributions is seen only for the gap-junctional network.
The other two networks exhibit exponentially decaying nature for both the
degree and the strength distributions.

{\bf Figure S3.} 
The intra- and inter-modular connectivity of individual neurons
in the C. elegans somatic nervous systems, color-coded to represent the
different ganglia in which each occurs.
(a) The within module degree $z$-score of each neuron in the empirical
neuronal network is shown against the corresponding participation
coefficient $P$.
(b) The corresponding result for a randomized version of the C. elegans
network where the degree of each neuron is kept unchanged.  The lateral
ganglion is seen to occupy a prominent position in the system, coordinating
the information flow between the neuronal groups responsible for receiving
sensory stimuli and those controlling motor activity. On the other hand,
the randomized network shows similar prominence for several other ganglia.

{\bf Table S1.} The classification of neurons according to their membership
in the 6 modules obtained by optimal partitioning of the combined
synaptic-gap junctional network of the C. elegans somatic nervous system.

{\bf Table S2.} The neuronal composition of different functional circuits
in the C. elegans somatic nervous system.

{\bf Table S3.} The classification of neurons in the C. elegans somatic
nervous system, according to their role in intra- and inter-modular
connectivity.

{\bf Text S1.} Analysis of the C. elegans network of synaptic connections.

\bibliography{/home/rajkp/Research/Drafts/Bibliography/modularity_network,celegans}
\end{document}